\documentclass[twocolumn,showpacs,aps,showkeys]{revtex4}
\usepackage{amsmath,bm}
\usepackage{graphicx}

\newcommand{\be}{\begin{equation}}
\newcommand{\ee}{\end{equation}}
\newcommand{\BE}{\begin{eqnarray}}
\newcommand{\EE}{\end{eqnarray}}
\newcommand{\BEn}{\begin{eqnarray*}}
\newcommand{\EEn}{\end{eqnarray*}}
\newcommand{\barr}{\begin{array}} 
\newcommand{\earr}{\end{array}}

\newcommand{\bit}{\begin{itemize}}
\newcommand{\eit}{\end{itemize}}
\newcommand{\bfl}{\begin{flusleft}}
\newcommand{\efl}{\end{flusleft}}
\newcommand{\bfr}{\begin{flushright}}
\newcommand{\efr}{\end{flushright}}
\newcommand{\bfig}{\begin{figure}}
\newcommand{\efig}{\end{figure}}

\newcommand{\bc}{\begin{center}}
\newcommand{\ec}{\end{center}}

\newcommand{\ben}{\begin{enumerate}}    
\newcommand{\een}{\end{enumerate}}

\newcommand{\eps}{\varepsilon}
\newcommand{\de}{\partial}

\begin{document}

\title{Analysis of PIN1 WW domain through a simple Statistical Mechanics Model}

\author{Pierpaolo Bruscolini}
\address{Instituto BIFI, Universidad de Zaragoza, 
c/o Corona de Arag\'on 42,
  E-50009 Zaragoza (Spain). }

\author{Fabio Cecconi}
\address{Dipartimento di Fisica Universit\`a "La Sapienza" \& INFM
Unit\`a di Roma1, P.le A.~Moro 2, I-00185 Roma.}

\begin{abstract}
We have applied a simple statistical-mechanics G$\bar{\text{o}}$-like model to the analysis
of the PIN1 WW domain, resorting to Mean Field and Monte Carlo techniques to
characterize its thermodynamics, and comparing the results with the wealth of
available experimental data. 
PIN1 WW domain is a 39-residues  protein fragment which folds on an
antiparallel $\beta$-sheet, thus representing an interesting model system to
study the behavior of these secondary structure elements.
Results show  that the model correctly reproduces the two-state behavior of
the protein, and also the trends of the experimental
$\phi_T$-values. Moreover, there is a good agreement between Monte Carlo
results and the Mean-Field ones, which can be obtained with a substantially
smaller computational effort. 
\end{abstract}



\maketitle

\section{Introduction}
Understanding the folding process of proteins is one of the most challenging
issues of biochemistry which requires sophisticated simulations at atomic
resolution generally referred as all-atom methods.
At present the large
incompatibility between folding time scales and regimes explored by all-atom
simulations makes the folding process not yet accessible to these powerful
computational approaches. 
Even though very encouraging progress have been achieved,
their applicability remains restricted to the
study of peptides and fragments of proteins {\cite{Karplus,Pande}.
In addition, the comparison to experiments
requires an accumulation of folding events to gain a enough large
statistics further narrowing the route to the
to full-atom techniques.  These limitations suggest resorting
to minimalist models which adopt a less accurate description of protein
chains, residue-residue and residue-solvent interactions
\cite{BW,HP,Shak,Guo,HPW}.
Approximate representations
reduce the computational costs and, with a certain amount of uncertainty,
allow to follow all the stages which bring a protein into its
native fold. The use of simplified models within a statistical mechanical
approach to protein folding is grounded on the assumption that not all the
chemical details need to be retained to understand and describe the basic
properties of folding processes. Of course the approximations, that this 
kind of approach introduces, must ensure that the basic principles of 
biochemistry are fulfilled to keep a correct description of the real 
molecules.  
Several years ago a simple model was
proposed by N. G$\bar{\text{o}}$ \cite{Go} to attain a
phenomenological but complete description of the folding reaction. 
The model replaces all 
non-bonded interactions by attractive
native-state contact energies. This recipe, which can be applied only when
native structure is known, implements the idea 
that a reasonable energy bias toward the native state could capture
the relevant features of the folding process.
This kind of modelling removes high energetic barriers along the pathways 
toward the native conformation (which lies in a deep minimum),  and produces 
relatively smooth energy landscapes. 
As a result the folding "funnel" \cite{Funnel1,Funnel2} leading to the 
native state is very smooth so the folding process results "ideal".
Folding events simulated through G$\bar{\text{o}}$-like potentials take 
only few nanoseconds  making possible to obtain statistically
meaningful results for generic proteins and polypeptide chains.
Since G$\bar{\text{o}}$-like models lack any energetic frustration, the scope
of their applications is related to the investigation of the role of
geometric frustration and configurational entropy in the folding process. 
Their success in providing a reasonable account for kinetic properties of the
folding process is related to the assumption that folding kinetics is mainly
determined by native geometry, together with native state stability, and
this view is indeed supported by several
experimental works \cite{OW2004,Chen,Plaxco,Santiago,Chiti}.
Along the lines indicated by the G$\bar{\text{o}}$-philosophy other
simplified models
exploiting the information present in the native state have been 
proposed \cite{AlmBak,MEpnas2,Sloop,Finkel}.   
In this paper we continue our analysis
\cite{BruCecc} of one of this
G$\bar{\text{o}}$-like models, the Finkelstein 
model \cite{Finkel,Finkel2004}, and apply
it to the study of the Pin1
WW domain (pdb code 1I6C) which has a well defined and
simple native structure made of two slightly bent antiparallel beta-sheets.
Its distinctive feature, which is also reflected in its name, is the presence 
of two Triptophanes (W), located 20-residues apart from one another.  
Its structure, with a simple topology,
lacks of all those features that can complicate the modeling.  Thus
this molecule represents a suitable candidate to explore the kinetic and
thermodynamic factors responsible for the formation of $\beta$-sheets and
their stability, and is also a suitable benchmark through which validate models
and theories.  
The Finkelstein model is particularly suitable for analyzing the
folding thermodynamics of two-state proteins and the WW domain is known to fold
in a two state scenario so we can test whether the model can faithfully
reproduce the known experimental data \cite{jmb2001} 
about WW domain folding.

The organization of the paper is as follow.
In section II we discuss the model and
its assumptions. In section III we present the Monte Carlo and Mean
Field methods we
adopt, and in section IV we report and discuss our results. Finally, 
section V is dedicated to the concluding remarks.     

\section{Description of Finkelstein Model}
Finkelstein model assumes a simple description of the polypeptide chain, where
residues can stay only in an ordered (native) or disordered
(non-native) state. Then, each micro-state of a protein with $L$
residues is encoded in a sequence of $L$ binary variables
${\bf s}= \{s_1,s_2,...,s_L\}$, $s_i=\{0,1\}$.
Residues with $s_i=1$ ($s_i=0$) are in their native (non-native)
conformation.
When all variables take on the value $1$, the protein is considered folded, 
whereas the random coil corresponds to all $0$'s.
Because each residue can be in one of the two states, ordered or disordered, 
the free energy landscape consists of $2^L$ configurations.
This enormous reduction in the number of configurations available
to a protein is a quite delicate point because  
it is a restrictive feature of the model. However this crude assumption,
already employed in \cite{Zwanzig}, is
the simplest one leading to a two state behaviour of the folding. 

The 
effective Hamiltonian (indeed, a free-energy function) is 
\begin{equation}
H({\bf s}) = \eps \sum_{i<j}  \Delta_{ij}  s_i s_j - T S({\bf s})\,,
\label{eq:finkel}
\end{equation}
where $S({\bf s})$ is given by:
\begin{equation}
S({\bf s}) = R \left[ q \sum_{i=1}^{L} (1 - s_i) + S_{loop}({\bf s})\right]\,.
\label{eq:S}
\end{equation}
$R$ is the gas constant 
and $T$ the
absolute temperature. 
The first term  in Eq.~(\ref{eq:finkel}) 
is the energy associated to native contact
formation. Non native interactions are neglected:  this further
assumption can be just tested {\it a posteriori} and it
is expected to hold if, during the folding process, the progress
along the reaction coordinate is well depicted on the basis of 
the native contacts. That is, the reaction coordinate(s) must be
related to just the native contacts. Moreover, such progress must be slow  
with respect to all other motions, so that all non-native interaction
can be ``averaged-out'' when considering the folding pathways.
$\Delta_{ij}$ denotes the element $i$,$j$ of the contact matrix,
whose entries are the number of heavy-atom contacts between 
residues $i$ and $j$ in the native state.
Here we consider two amino-acids to be in contact
if there are at least two heavy 
atoms (one from aminoacids $i$ and one from $j$) separated by a distance 
less than $5$\AA. 
The matrix $\Delta$ embodies the geometrical properties of the protein.

The second  term in Eq.~(\ref{eq:finkel}) is the conformational entropy
associated to the  presence of unfolded regions along the chain,
and vanishes in the native state.
 
More precisely the first term in Eq.~(\ref{eq:S}) is 
a sort of ``internal'' entropy of the residues: $q R$ represents
the entropic difference between the coil and the native
state of a single residue. This can be noticed by considering that in 
the fully unfolded state $S_{loop}$ vanishes and the remaining entropy is
$q L R$ only.

The term  $R S_{loop}$ in Eq.~(\ref{eq:S}) 
is the entropy pertaining
to the disordered closed loops protruding from the globular native 
state~\cite{Sloop};
it reads:
\begin{equation}
S_{loop}({\bf s}) = \sum_{i<j} J(r_{ij}) \prod_{k=i+1}^{j-1}(1-s_k)
s_i s_j\,.
\label{Eq:Sloop}
\end{equation}
According to \cite{Finkel}, we take:
\begin{equation}
J(r_{ij}) = -\frac{5}{2} \ln|i-j| -
\frac{3}{4}\frac{r_{ij}^2-d^2}{Ad|i-j|}\,.
\label{Eq:J}
\end{equation}
In this way the configuration of a disordered loop going from residues
$(i+1)$ to $(j-1)$, with $i$ and $j$ in their native positions,
is assimilated to a random walk
with end to end distance $r_{ij}$, the latter being the distance between
C$_{\alpha}$ atoms of residues $i$ and $j$ in the native state. 
The parameters $d=3.8$ \AA~ and $A=20$ \AA~ are the average distance 
of consecutive $C_{\alpha}$ along the chain and persistence length 
respectively.
The entropy of one loop closure~(\ref{Eq:J}) differs from the classical result 
$-3R/2 \ln(N)$  pertaining to a free Gaussian chains \cite{JS}. 
The presence of the factor $5/2$, instead of $3/2$, stems from the fact that a loop 
exiting the globule must  lie completely outside of it, to account for
the self-avoidance. Thus, the spatial domain occupied by the globule results
in a forbidden region for the 
disordered loop, and  
this simple sterical constraint, reducing the number of accessible 
conformations, 
increases the entropy loss obtained from the closure
of the loop \cite{Sloop}.        
\section{Methods}
A direct comparison between model predictions and experimental results
requires a tuning of the coefficients $q$ and $\varepsilon$ in the
energy function Eq.~(\ref{eq:finkel}).
In our computation we set $q=2.31$ and regarded $\varepsilon$ as an
adjustable parameter. We determined it by imposing that
the mean-field specific heat exhibits its ``collapse'' peak 
in correspondence to the experimental transition temperature $T=332$ K~
\cite{jmb2001}. 
Despite the use of a simple MF approach, we expect that this procedure yields
a correct estimate for $\varepsilon$, since the MF is known to
reproduce  the thermodynamics properties of the Finkelstein model
pretty faithfully~\cite{BruCecc}.
Once determined the optimal choice of $q$ and $\varepsilon$, we performed
Monte Carlo simulations to investigate the thermal folding of the WW domain.
We implemented a Metropolis algorithm
with transition rates between states $j$ and $k$
$$
w(j\to k) = \exp[(H_j - H_k)/RT]
$$ 
$R$ being the gas constant, $T$ the temperature and $H_j$ the 
Finkelstein energy of state $j$, according to Eq.~(\ref{eq:finkel}). 

We applied the multiple histogram technique (MHT) \cite{Ferren} to
reconstruct the system density of states (DOS) in the full range
of accessible energies. To this end, we carried out MC runs at $50$
equally spaced temperatures in the range $273-383$ K, and for each run
we collected the energy histogram to estimate the statistical weight of all
configurations with a certain energy.
Through the Swendsen-Ferremberg
procedure \cite{Ferren} these histograms were optimally linearly combined
to extract the whole DOS and thus compute the entropy
$S(E) = R\ln[g(E)]$ up to an additive constant. 
The knowledge of entropy allows evaluating the
free energy profiles $F(E) = E - TS(E)$, and
other relevant thermodynamical quantities for the folding, such as
the specific heat.

In its variational formulation \cite{VarMF}, 
Mean Field Approximation, for a system with Hamiltonian $H$ and corresponding  
free-energy $F$, amounts to minimizing
\begin{equation}
F_{var} \leq F_0 + \langle H - H_0 \rangle_0\,,
\label{Eq:genericFvar}
\end{equation}
where $H_0$ is a solvable trial Hamiltonian $F_0$ is the corresponding
free-energy, both depending on free parameters
${\bf x} = \{x_1\cdots x_L\}$ (variational parameters). 
Minimization leads to the self consistent equations
that in their general form read
\begin{equation}
\bigg\langle \frac{\partial H_0}{\partial x_l} \bigg\rangle_0 
\langle H-H_0\rangle_0 - \bigg \langle (H-H_0)
\frac{\partial H_0}{\partial x_l}
\bigg\rangle_0 = 0\,, 
\label{Eq:meanfield}
\end{equation}
 with $l=1,\ldots, L$. We have implemented different versions of the MFA for
the model that differ each from the other 
by the choice of the trial Hamiltonian.

The standard MFA employees as the
trial Hamiltonian: 
\begin{equation}
H_0 = \sum_{i=1}^L x_i s_i \,,
\label{eq:H_0mfa1}
\end{equation} 
with $x_i$ to
be determined by minimizing the variational free-energy \cite{VarMF}
\begin{equation}
F_{var}({\bf x},T) = \sum_{i=1}^L f_0(x_i,T) + 
\langle H - H_0 \rangle_0 \,, 
\end{equation}
where $\sum_i f_0(x_i,T)$ is the free energy associated to $H_0$,
\begin{equation}
f_0(x_i,T) = -\frac{1}{\beta} \ln \bigg\{1 + \exp(-\beta x_i)\bigg\}\;.
\end{equation}  
Thermal averages, performed through the Hamiltonian $H_0$,
factorize 
$\langle s_i s_j ... s_k \rangle_0 = 
\langle s_i \rangle_0 \langle s_j \rangle_0 ... \langle s_k \rangle_0$.    
The approximate average site ``magnetization''
$m_i = \langle s_i \rangle_0$  depends only on the field $x_i$,
and is given by
\begin{equation}
m_i = \frac{\de F_0}{\de x_i} = \frac{1}{1 + \exp(\beta x_i)}\,.
\label{eq:m_i}
\end{equation}
Instead of working with external fields $x_i$'s, it is more intuitive
to use the corresponding ``magnetizations'' $m_i$'s, writing $F_{var}$
as a function of the $m_i$'s.
Due to the choice of $H_0$, Eq.~(\ref{eq:H_0mfa1}), and to the
expression Eq.~(\ref{eq:m_i}),
evaluating  the thermal average $\langle H \rangle_0$ amounts to 
replacing, in the Hamiltonian Eq.~(\ref{eq:finkel}),  
each variable $s_i$ by its thermal
average $m_i$.
In the end we get:
\BE
F_{var}({\bf m},T) &=& \eps \sum_{ij} \Delta_{ij} m_i m_j -
T S({\bf m})  \nonumber \\
&& +R T \sum_{i=1}^L g(m_i)\,,
\label{eq:Fmvar}
\EE
where $g(u) = u \ln(u) + (1-u) \ln(1-u)$ and $S({\bf m})$ is obtained
from Eq.~(\ref{eq:S}) by substituting $s_i \rightarrow m_i$. 
The last term corresponds to 
$ F_0 - \langle H_0 \rangle_0$ in Eq.~(\ref{Eq:genericFvar}):
it is the entropy
associated to the system with Hamiltonian $H_0$ and is the typical
term that stems from this kind of MFA \cite{VarMF}. 
The minimization of function Eq.~(\ref{eq:Fmvar})
with respect to ${\bf m}$ leads to 
self-consistent equations:
\begin{equation}
g'(m_i) =  \eps \sum_{j} \Delta_{ij} m_j -
R T \bigg(q - \frac{\de S_{loop}({\bf m}) }{\de m_i}\bigg)\,.
\label{eq:self}
\end{equation}
Equations~(\ref{eq:self}) can be solved numerically
by iteration and provide the
optimal values of the magnetizations that we denote by ${\bf m}^*$.
Once the set of solutions ${\bf m}^*$ is available, we can compute the 
variational free-energy  $F_{var}({\bf m}^*)$
that represents the better estimate of the system free-energy $F$.
Free energy profiles are evaluated performing the minimization after the
introduction of Lagrange multipliers, corresponding to the constraint of
considering states with a fixed number of native residues.

A different MFA consists in taking a trial Hamiltonian  that accounts exactly
for the entropic term of the original one, resorting to the procedure
introduced in \cite{Brusco},
and approximates the interactions
by introducing a weight dependent on the number of native residues in the
configuration. 
Namely, we consider the set of configurations of the proteins with $M$
native residues  ($M=0,...,L$) 
and take as the trial Hamiltonian
\begin{equation}
H_0({\bf x}) = \sum_{M=0}^L \delta(M - \Sigma_{i} s_i)
H_0^{(M)}({\bf x}) \,,
\label{Eq:H0difficult}
\end{equation}
where $\delta(\bullet)$ is the Kronecker delta,  and
$H_0^{(M)}$ is the Hamiltonian restricted  to the configurations
with $M$ natives: 
\begin{equation}
H_0^{(M)}({\bf x})= \sum_{i=1}^L \tilde{\varepsilon_i}\, x_i
\frac{M-1}{L-1} s_i - T S({\bf s}) \,, 
\label{Eq:H0M}
\end{equation}
with   $\tilde{\varepsilon_i} = (1/2) \sum_{j=1}^N \varepsilon_{i,j}
\Delta_{i,j}$.
Each  residue $i$, in a generic configuration with
$M$ native residues, feels an interaction $\tilde{\varepsilon_i}$
which it would feel in the native state, weakened by a factor $(M-1)/(L-1)$
(accounting for the fact that not all the residues are native),
times  the external field $x_i$, to be fixed by the mean field procedure.

The mean-field equations for this case can be found in Ref.~\cite{BruCecc}. 

\section{Results and discussion}
The folding transition is signalled by the behavior of the specific heat, 
which develops a peak identifying the $T_f$. Standard MF peak position is
imposed to the 
correct experimental folding temperature to fit the parameters; notice though
that MC peak is correctly found at the same position, providing a consistency
check between the two methods (Fig.~\ref{fig:compare}).  

\begin{figure}
\includegraphics[clip=true,keepaspectratio,width=8.cm]{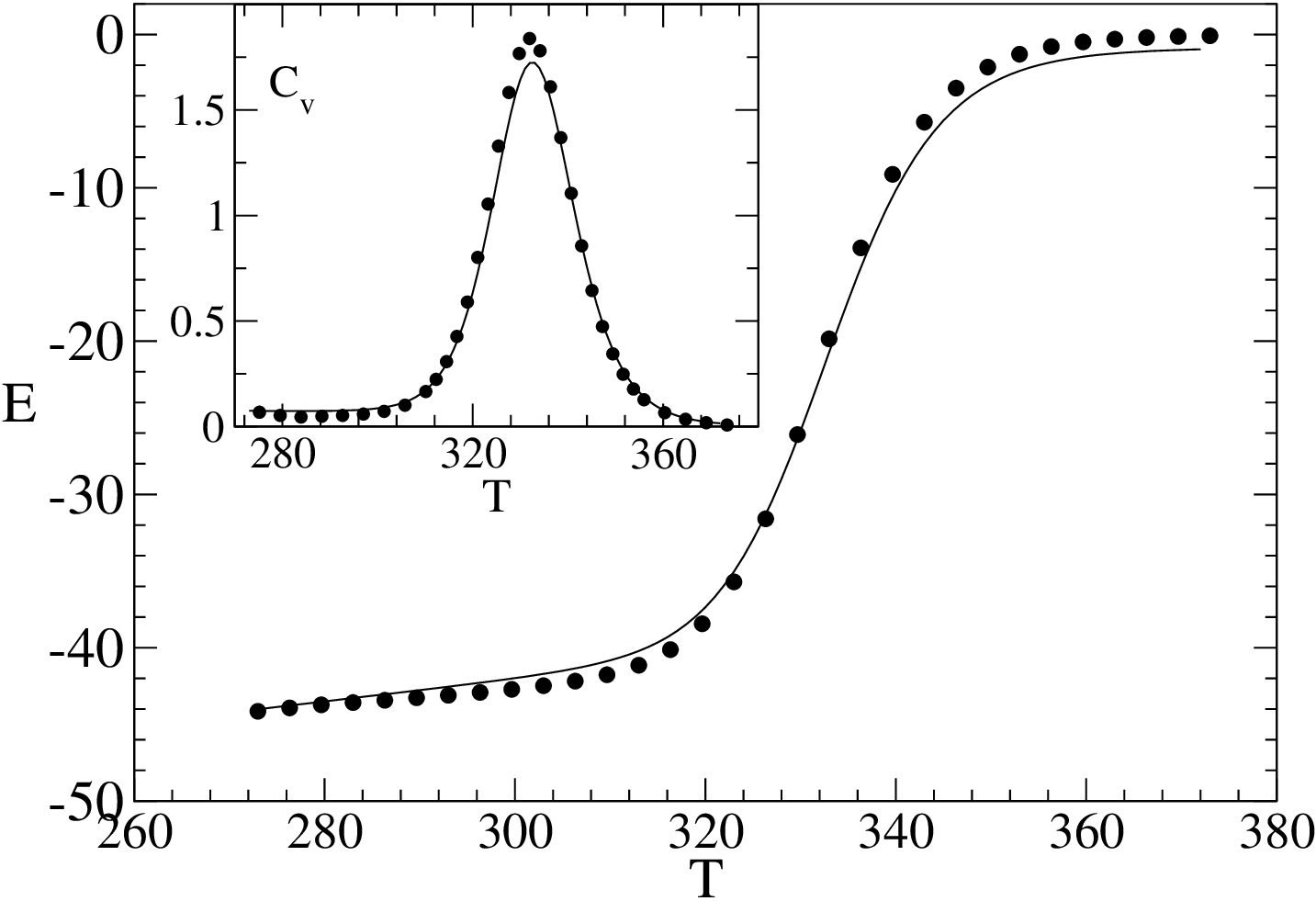}
\caption{Specific heat in Kcal mol$^{-1}$ T$^{-1}$ (inset) and energy (in Kcal
  mol$^{-1}$) as function of temperature, computed through MC simulations 
  (points) and standard Mean Field Approach (line).}
\label{fig:compare}
\end{figure}

Pin1 WW domain is reported to be a two-state folder~\cite{jmb2001}: this 
is recovered
by both the MC and the MF approximations, as can be seen in
Fig.~\ref{fig:probnat}. MC and the more complicated MF approach reproduce
with reasonable accuracy the experimental signal.

\begin{figure}
\includegraphics[clip=true,keepaspectratio,width=8.cm]{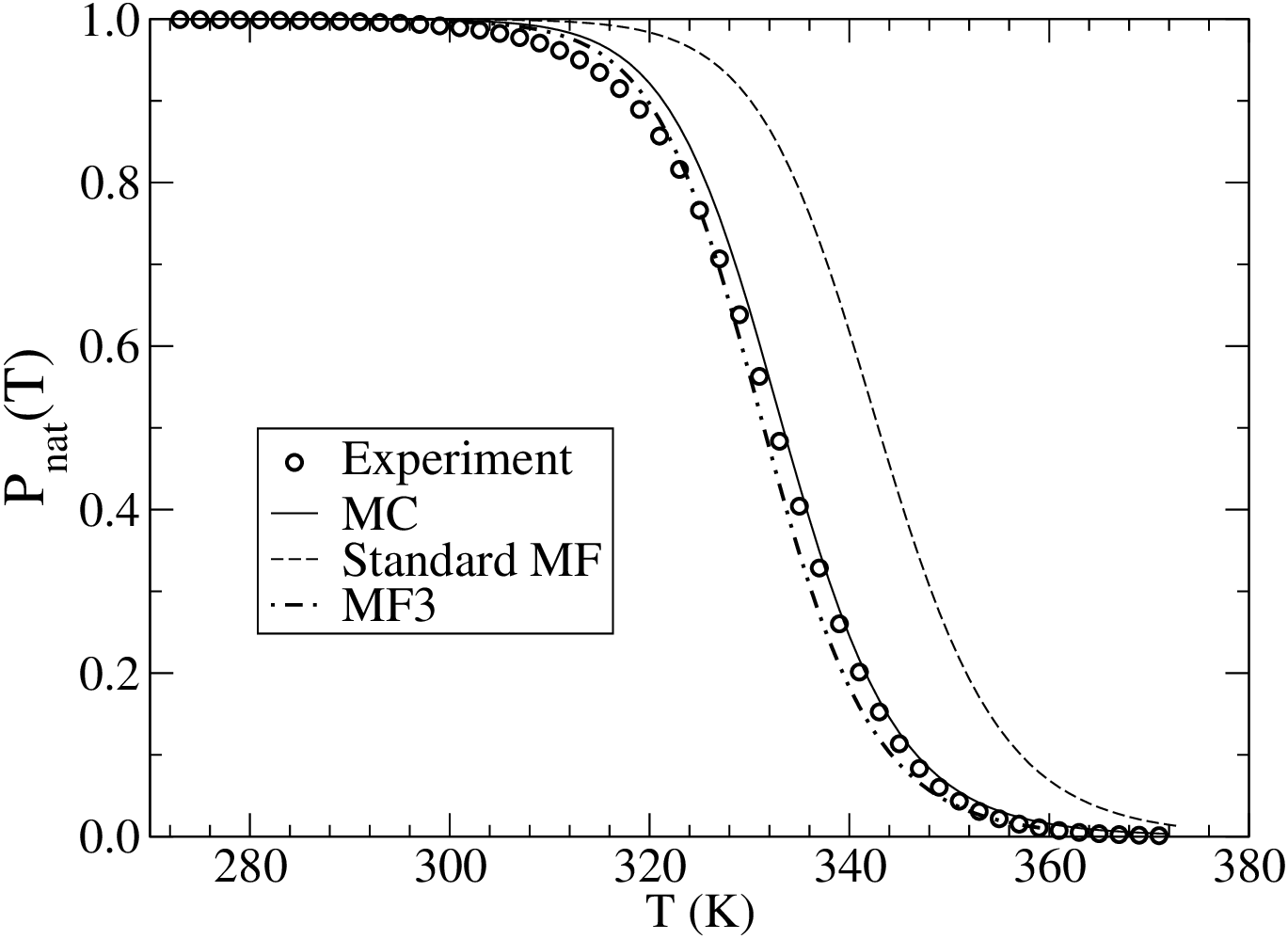}
\caption{Fraction of native protein as a function of temperature: MC
  simulation, standard Mean Field, Mean Field 3 of
  Ref.~\protect{\cite{BruCecc}} compared with
  the experimental fit in Ref.~\cite{jmb2001}}
\label{fig:probnat}
\end{figure}

The two-state nature of the protein can also be seen 
in the free-energy profiles
Figs.~\ref{fig:profMC},\ref{fig:profMF}. It is remarkable that the barrier
separating folded from unfolded conformations is quite
flat, especially in the MC case, so that mutations could likely
induce relevant changes in its position with just a slight change in the
energies, a scenario which is indeed suggested in
Ref.~\cite{jmb2001}. 

\begin{figure}
\includegraphics[clip=true,keepaspectratio,width=8.cm]{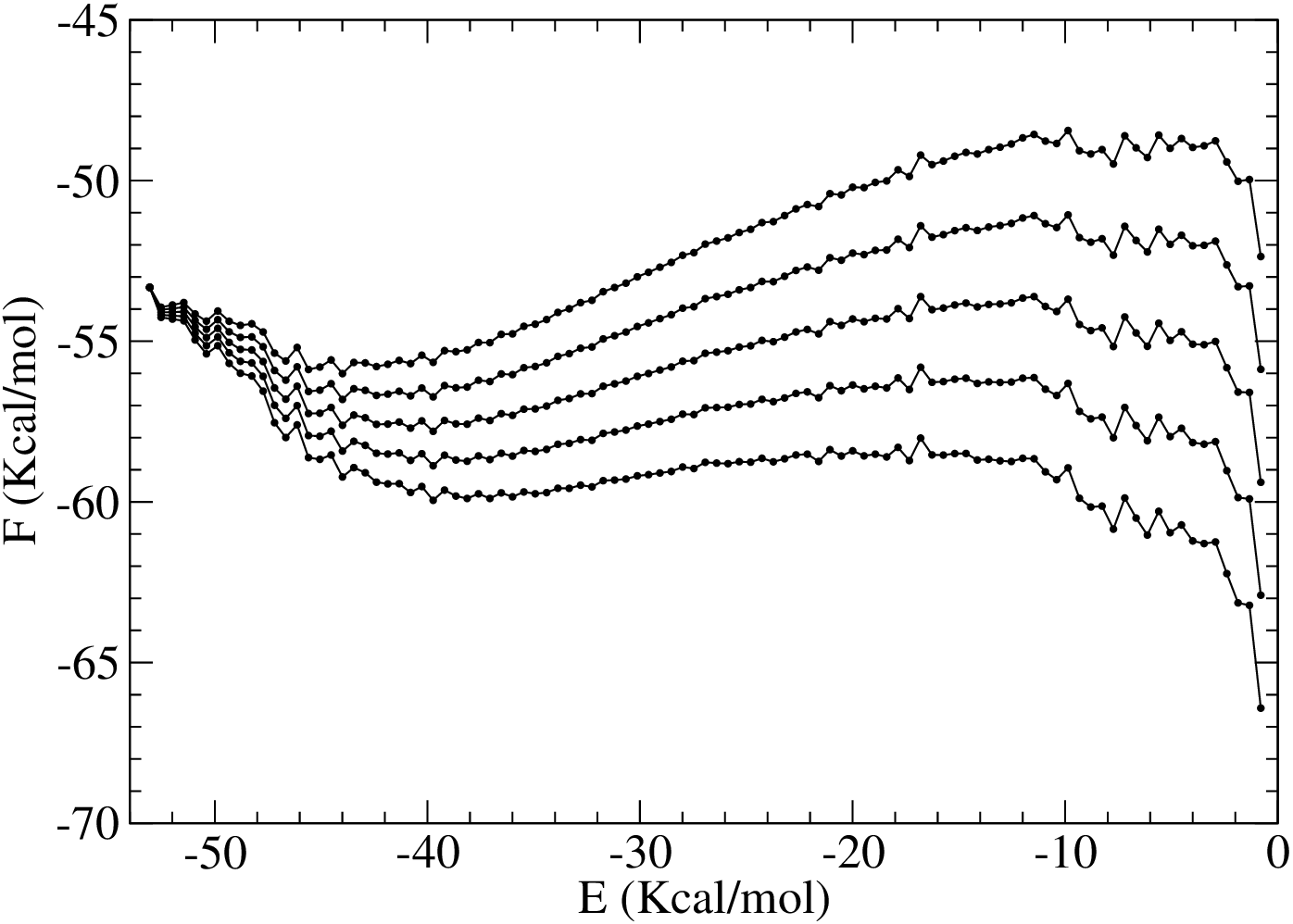}
\caption{MC free energy profiles, with the energy as the coordinate of
  reaction,  at different temperatures: from
  top to bottom $T$=292, 312, 332, 352, 372 K.}
\label{fig:profMC}
\end{figure}

\begin{figure}
\includegraphics[clip=true,keepaspectratio,width=8.cm]{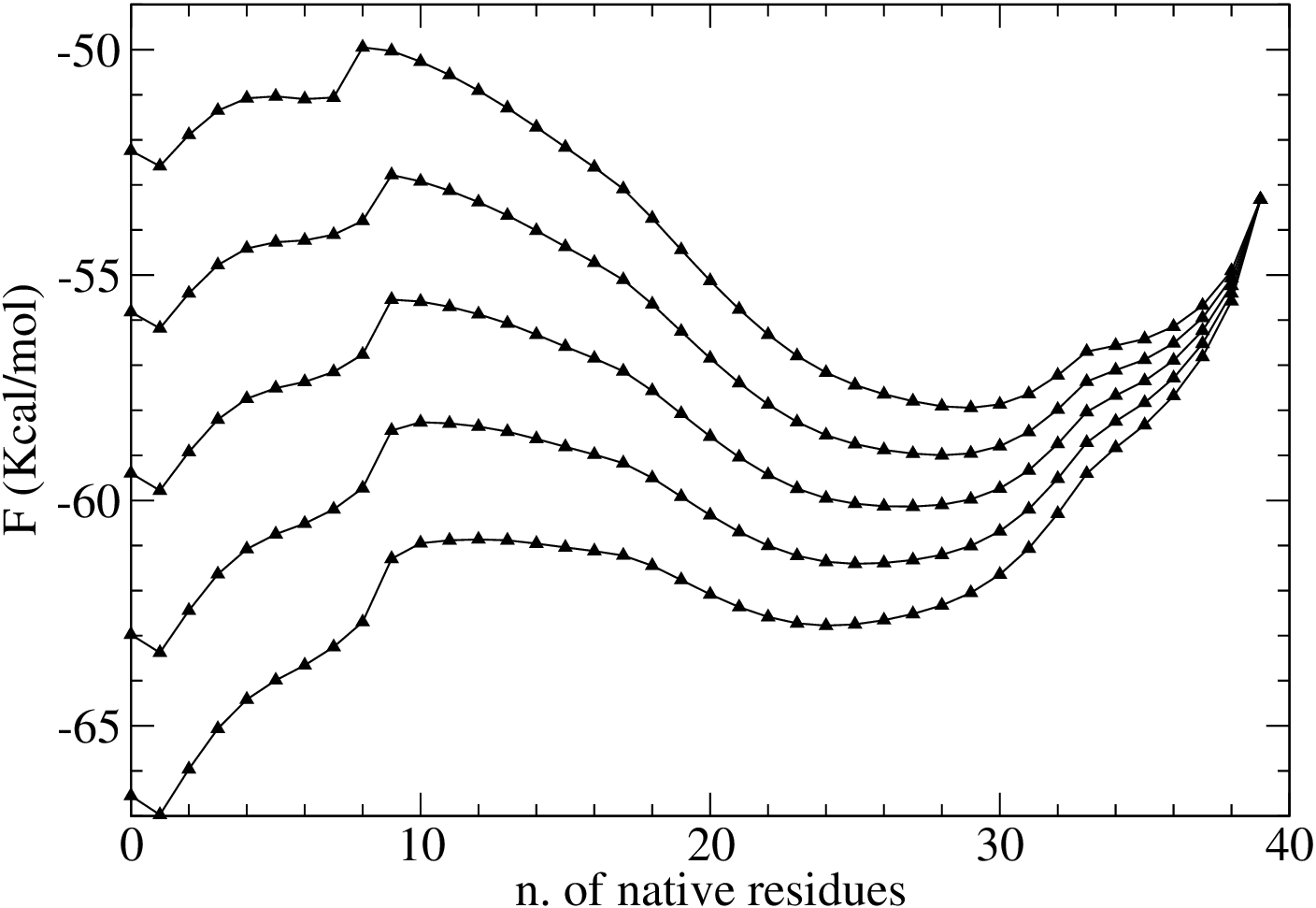}
\caption{Standard MF free energy profiles, in the number of native residues,  at different temperatures: from
  top to bottom $T$=292, 312, 332, 352, 372 K.}
\label{fig:profMF}
\end{figure}


Monte Carlo and Mean Field free energy profiles allow to estimate
the stability gap $\Delta G $ and the folding barrier $\Delta G^{\dag}$
as a function of temperature.
The comparison with the corresponding experimental
curves (Ref.~\cite{jmb2001})
\begin{eqnarray}
\Delta G_{ex}(T) = \Delta G_0 + \Delta G_1(T-T_f) + \Delta G_2(T-T_f)^2
\nonumber \\
\Delta G^{\dag}_{ex}(T) = \Delta G^{\dag}_0 + \Delta G^{\dag}_1(T-T_f)
+ \Delta G^{\dag}_2(T-T_f)^2  \nonumber
\end{eqnarray}
where $T_f=332$ K, 
$\Delta G_{0,1,2}$ = $\{-0.062,0.105,6.244 \cdot 10^{-4}\}$
Kcal/mol and 
$\Delta G^{\dag}_{0,1,2}$
=$\{5.089,0.0568,1.232 \cdot 10^{-3}\}$ Kcal/mol.
The result of this comparison is reported in Fig.~\ref{fig:DG}. 

\begin{figure}
\includegraphics[clip=true,keepaspectratio,width=8.cm]{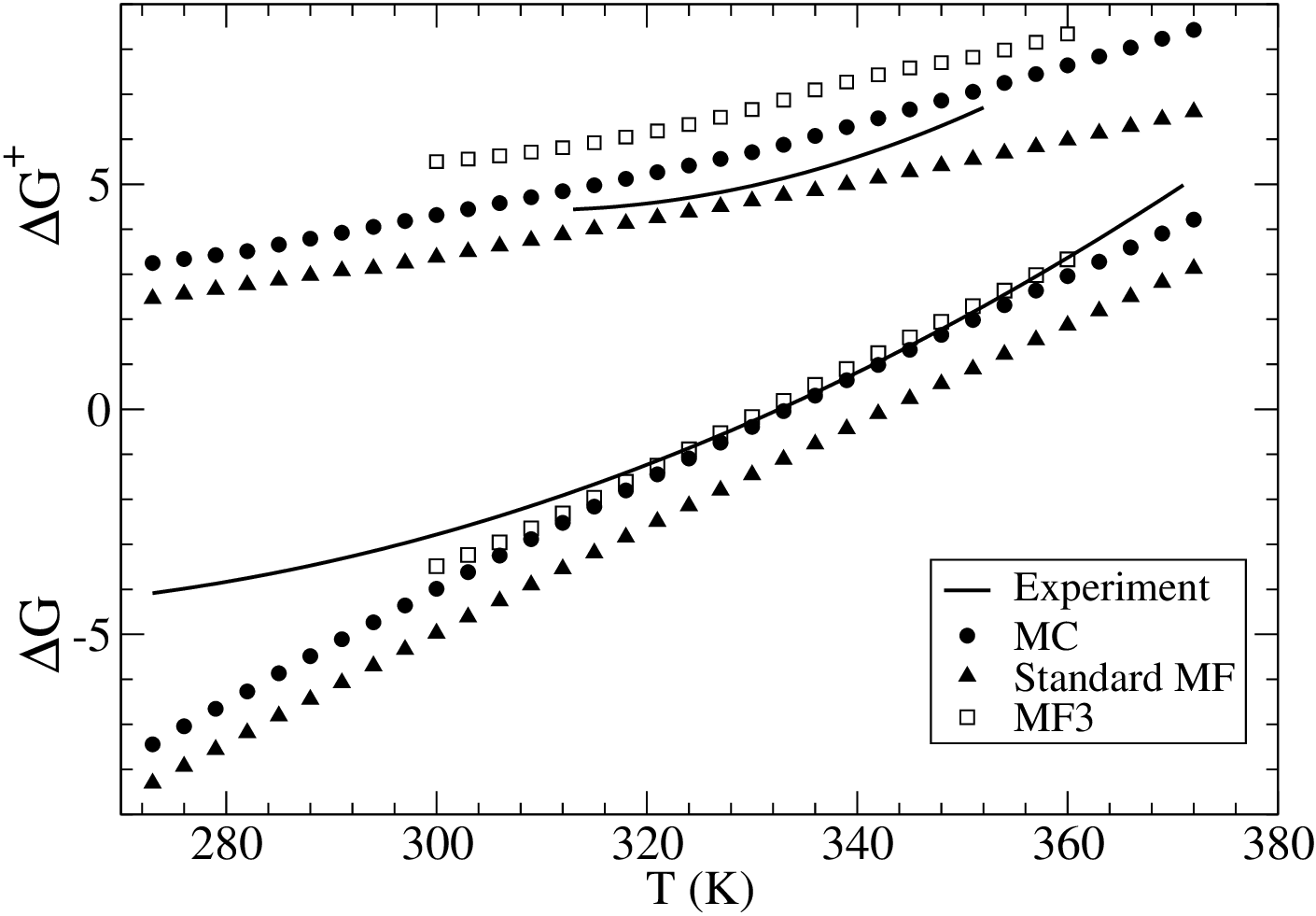}
\caption{Folding barrier (top set of curves) and stability of the native 
state
(bottom set) as a function of T, from experiment and simulations. 
Data are reported in Kcal/mol. 
}
\label{fig:DG}
\end{figure}
Notice that all methods compare
most favorably with the experimental results in the vicinity of $T_f$,
which is to be expected, since the model only accounts for the geometry, and
not for the details of the interactions, with their temperature dependence in
the hydrophobic contributions. MC gives a good estimate of both the stability
gap and the barrier, while standard mean field gives a reasonable description
of the folding barrier, but overestimates the stability. On the other hand, the
more complicated MF scheme recovers correctly the stability, but it
overestimates the barrier, at least if  we 
consider, as we did in Ref.~\cite{BruCecc}, 
just the profile of $F_0$ (relying on the good approximation that
$F_0$ provides to $F_{var}$), without resorting to
the more correct, but computationally expensive minimization of a
constrained $F_{var}$.
A more accurate analysis of free energy profiles
within this MF scheme is left for  future work. 
In the following, we analize standard MF and MC results concerning 
another important experimental quantity, namely the $\phi_T$-values
(Fig.~\ref{fig:phiT}). 
$\phi_T$-values are defined as 
\begin{equation}
\phi_T= \frac{\partial \Delta G^\dagger}{\partial T} \frac{1}{\frac{\partial
    \Delta G}{\partial T}} = \frac{\Delta S^\dagger}{\Delta S} \,\,\,,
\end{equation}
and give an idea of the entropy of the barrier compared to that of
the native state, providing a measure of the proximity of the barrier to
the folded state. 
The  experimental results show a monotonically increasing, continuous function,
spanning a wide range of values. MC and MF results indeed agree in the
monotonically increasing behavior, reflecting thus the Hammond behavior
~\cite{Hammond,Hammond2}, even if in a
discretized version. Indeed they
show a series of discrete jumps that, in the case of MC simulations, are not
simply an effect of the binning in the reaction
coordinate, but seem to suggest sharp movements in the barrier position:
sudden changes in   $\phi_T$ are in complete correspondence to shifts
in the position of the barrier, as reported in Fig.~\ref{fig:phiT}.

\begin{figure}
\includegraphics[clip=true,keepaspectratio,width=8.cm]{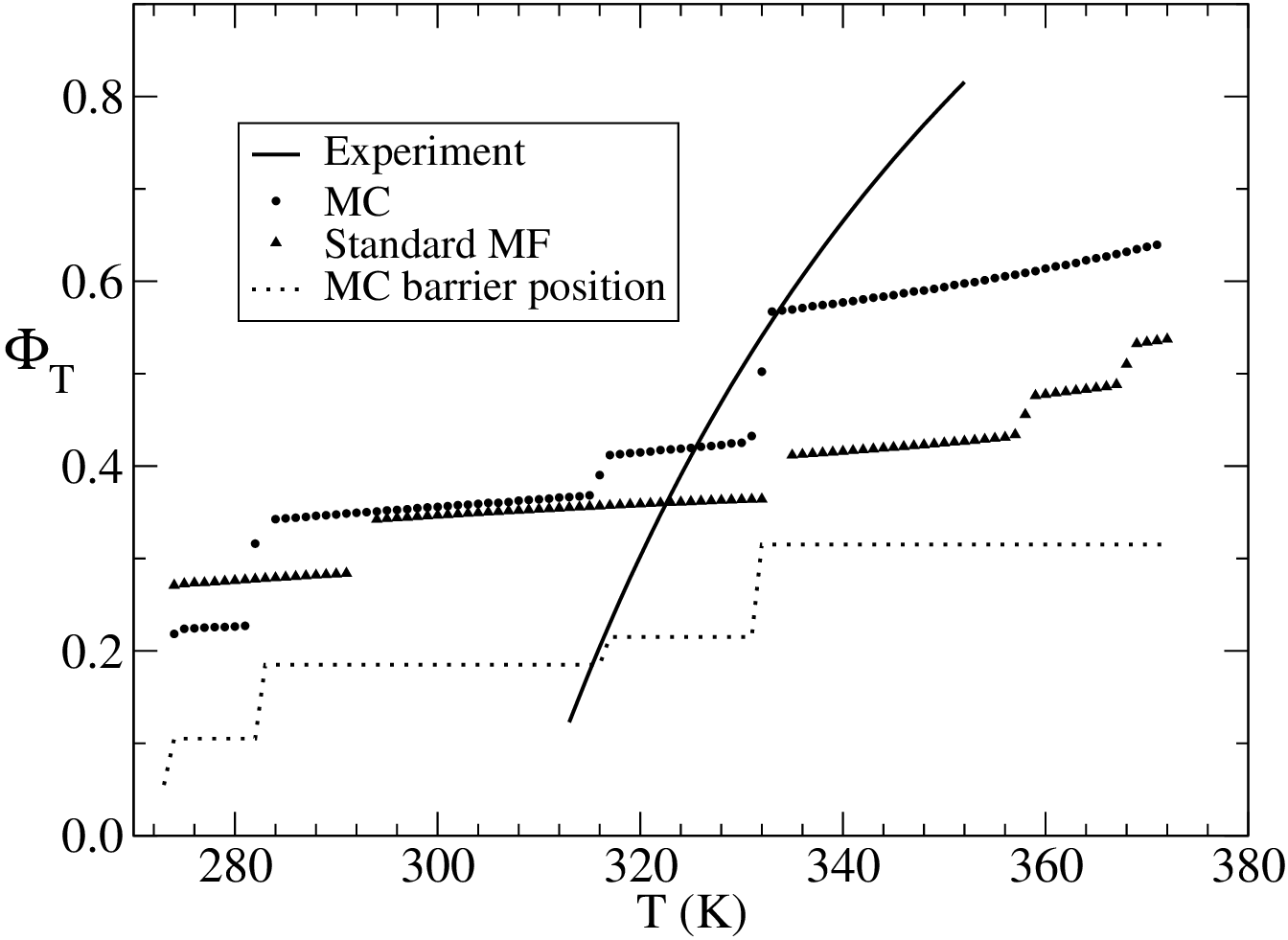}
\caption{$\phi_T$-values from experiments and simulations, together with the
  barrier position for the MC case. Barrier position values at a give T are
  evaluated as 
  the energy coordinate (x-axis in Fig.~\protect{\ref{fig:profMC}})
  corresponding  to the barrier top at that temperature,
  normalized to the total contact energy in
  the native state (independent from 
  T: $E=-53.32$ Kcal/mol with our choice of the 
  parameters). Notice how the shifts in $\phi_T$ correspond to those in the
  barrier position.}
\label{fig:phiT}
\end{figure}

\section{Conclusions}
The application of the Finkelstein model to protein PIN1 WW domain reveals
that this model, after fitting the parameter $\varepsilon$ in order to
reproduce the  correct transition temperature, is able to
describe correctly the thermodynamics of the folding process, at least in the
case of  simple two-state behavior. Indeed, the estimate of the folding
barrier, both in the case of MF approximation as well as for MC simulations,
lies within a relative  error of about 15\%  from the experimental estimate in
all the region of experimental measures. This is
indeed interesting, as the model lacks every detail about the nature of the
residues, dealing with all atomic contacts in the ground-state on the same
footing. Moreover, the estimate of the entropy is based on the theory of
noninteracting polymers, and neglects possible clashes of the protruding
unfolded loops with the folded part of the protein. 

Another important result concerns the $\phi_T$-values: both MF and MC results
recover the non-decreasing nature of experimental values, with MC providing a
better estimate of the slope than MF. At difference with the experimental
values, though, theoretical $\phi_T$-values increase in a discontinuous
fashion, with abrupt changes followed by steady plateaus. This behavior is
related to the fact the the transition state is quite broad, so that the
actual free-energy maximum, determining the barrier, jumps through different
values of the reaction coordinate (the number of native residues or the
energy). This is an aspect that deserve further analysis, also because the
simple three-state model, with a negligible intermediate, put forward by the
author of Ref.~\cite{jmb2001} does not seem to be able to reproduce the
experimental results with sufficient accuracy, and  a satisfactory
description of the transition state of this protein has still to be found. 
Probably, it  will require the introduction of residue heterogeneities and more
accurate studies on the dynamics of the system.


\end{document}